\documentclass[twocolumn,floatfix,aps]{revtex4-2}
\usepackage{bm}
\usepackage{mathbbol}
\usepackage{graphicx,color}
\usepackage[dvipsnames]{xcolor}
\usepackage{amsmath,amssymb}

\begin{document}
\title{Bifurcation theory captures band formation in the Vicsek model of flock formation} 
\author{C. Trenado} 
\affiliation{Department of Chemical and Biological Engineering, Princeton University, Princeton, NJ, 08544, USA}
\affiliation{Lewis-Sigler Institute for Integrative Genomics, Princeton University, Princeton, NJ 08544, USA}
\author{L. L. Bonilla$^{*}$}
\affiliation{Department of Mathematics, Universidad Carlos III de Madrid, 28911 Legan\'es, Spain}
\affiliation{G. Mill\'an Institute for Fluid Dynamics, Nanoscience and Industrial Mathematics, Universidad Carlos III de Madrid, 28911 Legan\'es, Spain\\
$^*$Corresponding author. E-mail: bonilla@ing.uc3m.es}
\author{A. Marquina}
\affiliation{Departamento de Matem\'atica Aplicada, Universitat de Val\`encia,
Dr. Moliner 50, 46100, Burjassot (Val\`encia), Spain} 

\date{\today}
\begin{abstract}
Collective behavior occurs ubiquitously in nature and it plays a key role in bacterial colonies, mammalian cells or flocks of birds. Here, we examine the average density and velocity of self-propelled particles, which are described by a system of partial differential equations near the flocking transition of the Vicsek model. This agent-based model illustrates the trend towards flock formation of animals that align their velocities to an average of those of their neighbors. Near the flocking transition, particle density and velocity obey partial differential equations that include a parameter $\epsilon$ measuring the distance to the bifurcation point. We have obtained analytically the Riemann invariants in one and two spatial dimensions for the hyperbolic ($\epsilon=0$) and parabolic ($\epsilon\neq 0$) system and, under periodic initial-boundary value conditions, we show  that the solutions include wave trains. Additionally, we have found wave trains having oscillation frequencies that agree with those predicted by a linearization approximation and that may propagate at angles depending on the initial condition. The wave amplitudes increase with time for the hyperbolic system but are stabilized to finite values for the parabolic system. To integrate  the partial differential equations, we design a basic numerical scheme which is first order in time and space. To mitigate numerical dissipation and ensure good resolution of the wave features, we also use a high order accurate WENO5 reconstruction procedure in space and a third order accurate Runge-Kutta scheme in time. Comparisons with direct simulations of the Vicsek model confirm these predictions.
\end{abstract}


\maketitle

\section{Introduction} \label{sec:1}
Biological systems exhibit self-organized collective motion across scales, from bacterial communities \cite{bud91,bud95,drescher11}, tissue development \cite{trepat09}, and cancer progression \cite{friedl09,aber62} to swarms of insects \cite{anstey09} or flocks of birds \cite{cav10,cav15}. Within these coordinate motions, traveling patterns arise and a variety of examples include patterns of waves produced by E. coli bacteria \cite{Hong20}, oscillating chemical reactions \cite{field85} and different processes in biology and medicine \cite{mur03}. Most mathematical models that capture such spatiotemporal dynamics are based on numerical simulations of agent-based models. Here, we examine a minimal discrete model of self-propelled particles called the Vicsek model (VM)  \cite{vic95}. It encompases the paradigm of dry active matter \cite{mar13}  whose components influence their collective motion without being immersed in a fluid or other medium. In the VM, $N$ particles move with equal speed in a box with periodic boundary conditions, and their positions are updated in fixed times so that the velocity of each particle adopts the direction of the average velocity of its close neighbors with some alignment noise ({\em conformist or majority rule}) \cite{vic95,vic12}. For different variants and applications of the VM, see \cite{mar13,ram10,vic12,mig18,ihl11,cho12,ihl15,ihl16,BT18,bon19}. The VM exhibits a known phase transition from disordered to coherent behavior that is an example of spontaneous symmetry breaking out of equilibrium. For the VM with forward update and a box below a critical size, decreasing alignment noise or increasing particle density makes the system go through a continuous {\em flocking} phase transition from a disordered state with uniform density to an ordered state characterized by nonzero average speed of the particles \cite{vic95}. For box size larger than critical, the flocking transition is discontinuous and a variety of patterns are possible \cite{gre04,cha08,sol15,hue08}. However, for a VM with backward update, the phase transition is continuous independently of the box size \cite{bag09}. 

While most of the results on the VM are based on direct numerical simulations, there exist a number of studies based on kinetic and hydrodynamic theories \cite{mar13}. Early kinetic theories are based on a Boltzmann transport equation \cite{ber06} and an Enskog equation with discrete time and continuous space \cite{ihl11}. In both cases, the authors aim to derive hydrodynamic equations proposed by Toner and Tu on the basis of symmetry considerations \cite{ton95,ton05}. These successful derivations are based on closing a hierarchy of equations for the Fourier coefficients of the distribution function and scaling assumptions \cite{ber06,ihl11,ihl16}. 

For the discrete time Enskog equation, two of the present authors have studied the stability of the disordered uniform density state and the corresponding bifurcation theory \cite{BT18,bon19}. The bifurcation solutions issuing from the disordered state depend on a small amplitude parameter $\epsilon$ that scales as the square root of the distance of the bifurcation parameter (e.g., alignment noise) to its critical value. The bifurcating states satisfy partial differential equations (PDEs) for slowly varying time and space scales. They still contain the small amplitude parameter $\epsilon$ and their character depends on it: the bifurcation equations are hyperbolic for $\epsilon=0$ and parabolic for $\epsilon\neq 0$ \cite{BT18}. On short time scales, the terms proportional to $\epsilon$ can be ignored and the resulting hyperbolic equations may exhibit solutions that have a nontrivial space dependence and oscillate in time. For sufficiently long times, the extra diffusive and source terms dampen the oscillations and make the solutions approach standing waves. These phenomena will be explored in the present paper.

Despite the multiple numerical and theoretical studies of the VM, many questions are still under debate. The VM is known for having four differentiated phases: (i) ordered homogeneous phase, (ii) ordered phase with band formation, (iii) ordered cross sea phase and (iv) homogeneous phase without order. Here, we show that during the band formation and under periodic boundary conditions, the solutions of the bifurcation equations include wave trains. In addition, in this paper we analyze for the first time the oscillatory behavior of the solutions by means of a high order accurate finite difference scheme that makes conspicuous the hyperbolic effects of the PDEs under periodic initial-boundary values. There are two main advantages of this approach. On the one hand, we have obtained the analytic expressions of the Riemann invariants to design our basic first order scheme, ensuring an accurate and correct propagation of the wave train solution. On the other hand, we use a high order accurate reconstruction procedure in space and a third order accurate Runge-Kutta scheme in time. These procedures mitigate the numerical dissipation and provide a good resolution of the waves' local extrema in the approximate solution.

The rest of the paper is organized as follows. We review the VM, its formulation in terms of a kinetic equation for the distribution function, the uniform distribution, its linear stability and the leading order of a pitchfork bifurcation solution in Section \ref{sec:2}. Subsection \ref{sec:3} contains the amplitude equations for the bifurcating solutions. These equations are equivalent to a conservation law for a density disturbance and a parabolic equation for the current density. As a small parameter measuring the distance to the bifurcation point tends to zero, these equations become a hyperbolic system. Subsection~\ref{sec:3} indicates the uniform bifurcating solution and a simple oscillatory solution for the linearized hyperbolic system. Section \ref{sec:4} analyzes in more detail the hyperbolic system in one dimension (1D) including a discussion of the numerical schemes we use. Section~\ref{sec:5} considers the parabolic system in one and two dimensions. Section~\ref{sec:6} compares the results obtained in previous sections to those of direct simulations of the VM. Lastly, Section~\ref{sec:7}  contains our conclusions and the Appendices are devoted to more technical material. 

\section{Model and kinetic equation} \label{sec:2}
\subsection{Model}
In nondimensional units, the two dimensional (2D) VM consists of $N$ active particles moving with speed 1 in a square box of side $\ell$ with periodic boundary conditions. At time $t$, the position and velocity of particle $j$ ($j=1,\ldots, N$) are  $\mathbf{x}_j(t)$ and $\mathbf{v}_j(t)= (\cos\theta_(t),\sin\theta_j(t))$, respectively. The particles tend to align their velocities to an average of those of their neighbors, so that at nondimensional time $t+1$ we have
\begin{eqnarray}
\theta_j(t+1)=\mbox{Arg}\!\left(\sum_{|\mathbf{x}_l-\mathbf{x}_j|<\mathbb{r}_0} e^{i\theta_l(t)}\right)\!+\xi_j(t), \label{eq1}\\
\mathbf{x}_j(t+1)=\mathbf{x}_j(t)+(\cos\theta_j(t+1),\sin\theta_j(t+1)).\label{eq2}
\end{eqnarray}
 In Eq.~\eqref{eq1}, neighbors are all particles inside a circle of radius $\mathbb{r}_0$ centered at particle $j$ and including it. This {\em metric} concept of neighborhood is appropriate for insect swarms \cite{att14}. The relation of our units to dimensional ones is given in row I of Table \ref{table1} \cite{BT18}. Alternative nondimensionalized units used elsewhere in the literature \cite{cha08,sol15} are given in row II. For example, if $v_0=0.5$ in nondimensionalization II, $\mathbb{r}_0/(v_0\tau)=2$, and the corresponding space grid size and density are twice and one fourth, respectively, of the corresponding magnitudes in nondimensionalization I. $\xi_j(t)$ are independent identically distributed (i.i.d.) random alignment noises, selected with equal probability in the interval $-\eta/2<\xi<\eta/2$ with $0<\eta\leq 2\pi$. The parameter $\eta$ measures the width of the alignment noise and can we thought of as a tolerance to failure in the alignment rule. 
 \begin{table}[ht]
\begin{center}\begin{tabular}{c|c|c|c|c|c|c|c|}
 \hline
&$\mathbf{x}$, $\mathbb{r}_0$, $\ell$& $\mathbf{v}$ & $t$&$\Delta x$ &$\rho_0$&$\theta$, $\xi$\\ \hline
I&$v_0\tau$ & $v_0$ & $\tau$&$v_0\tau$ &$\frac{N}{\ell^2}(v_0 \tau)^2$& --\\ 
II&$\mathbb{r_0}$ & $\frac{\mathbb{r}_0}{\tau}$ & $\tau$& $\frac{v_0\tau}{\mathbb{r}_0}$&$\frac{N}{\ell^2}\left(\mathbb{r}_0\right)^2$ & --\\ 
\hline
\end{tabular}
\end{center}
\caption{Units for nondimensionalizing the Vicsek model. In these units, $v_0=1$, $\mathbb{r}_0\neq 1$, for nondimensionalization I, and $\mathbb{r}_0=1$, $v_0\neq 1$, for nondimensionalization II. Relations between space grid size and density are also included.}
\label{table1}
\end{table}

To quantify whether particles flock, we use the overall average velocity as a complex order parameter 
\begin{eqnarray}
Z=W\, e^{i\Upsilon}=\frac{1}{N}\sum_{j=1}^N e^{i\theta_j}. \label{eq3}
\end{eqnarray}
The polarization amplitude $W=|Z|\in (0,1)$ indicates whether the particles form a flock ($W>0$) or not ($W=0$). $\Upsilon$ is the average phase of the particles. Increasing the average number of particles inside the region of influence, $M=N\pi \mathbb{r}_0^2/\ell^2$, favors flocking as more and more particles try to move together. Then there are a critical value of $M$ above which the polarization is $W>0$ and below which $W=0$ (in the limit as $N\to\infty$). The alignment noise $\eta$ also has a critical value. However, it is $W>0$ below and $W=0$ above threshold, because increasing $\eta$ damages the flocking effect of the alignment rule.

\subsection{Kinetic equation}
In the limit as the number of particles goes to infinity and the average particle density $\rho_0=N/\ell^2$ is sufficiently large, it is possible to derive a kinetic equation for the VM following Refs.~\cite{ihl11,ihl16,BT18,bon19}. The crucial assumption is that the $N$-particle distribution function equals the product of $N$ one-particle distribution functions \cite{ihl11}. The latter, $f(\mathbf{x},\theta,t)$, gives the number of particles at time $t$ having phases in $(\theta,\theta+d\theta)$ that are in a box of area $dx\, dy$ about $\mathbf{x}=(x,y)$. The kinetic equation for $f$ is discrete in time and space, it has a simple constant solution (the disordered state), and it is possible to work out its linear stability and bifurcation theory \cite{BT18,bon19}. The constant solution  $f_0=\rho_0/(2\pi)= N/(2\pi\ell^2)$ is linearly stable for noise $\eta>\eta_c$, where the critical noise satisfies \cite{ihl16}  
\begin{eqnarray}
 Q_1\sim\frac{\sqrt{\pi M}}{\eta}\sin\frac{\eta}{2}= 1\quad (M\gg 1), \quad M=\rho_0\pi\frac{\mathbb{r}_0^2}{\ell^2},\label{eq4}
\end{eqnarray}
where $M$ is the average number of neighbors in the circle of influence. Near the critical noise $\eta_c$, the disordered state undergoes a pitchfork bifurcation. The bifurcating solutions have the form \cite{BT18}
\begin{subequations}\label{eq5}
\begin{eqnarray}
&&f\sim f_0\!+\epsilon \frac{r(\textbf{X},T;\epsilon)+2\,\mathbf{w}(\textbf{X},T;\epsilon)\!\cdot\! (\cos\theta,\sin\theta)}{2\pi},\quad\label{eq5a}\\
&&\eta=\eta_c+\epsilon^2\eta_2 Q_\eta, \quad\textbf{X}=\epsilon\textbf{x},\quad T=\epsilon t.  \label{eq5b} 
\end{eqnarray}\end{subequations}
Here, $r(\textbf{X},T;\epsilon)$ is a disturbance of the average particle density $\rho_0=N/\ell^2$ and $\mathbf{w}(\textbf{X},T;\epsilon)$ is a particle current. The latter and Eq.~\eqref{eq5a} produce the complex order parameter of Eq.~\eqref{eq3}: 
\begin{eqnarray}
(\mbox{Re}Z,\mbox{Im}Z)\sim \epsilon\,\mathbf{w}(\mathbf{X},T),\quad W\sim\epsilon\,|\mathbf{w}(\mathbf{X},T)|.\label{eq6}
\end{eqnarray}

\subsection{Bifurcation equations}\label{sec:3}
The bifurcation equations are derived in Ref.~\cite{BT18} by using a Chapman-Enskog perturbation method that starts with  Eqs.~\eqref{eq5a}-\eqref{eq5b}. They are
\begin{subequations}\label{eq7}
\begin{eqnarray}
\frac{\partial r}{\partial T}\!&+&\!\nabla_X\!\cdot\!\left[\!\left(1+\frac{\epsilon\gamma_3r}{2\pi}\right)\!\mathbf{w}\right]\!=0,\label{eq7a}\\
\frac{\partial \mathbf{w}}{\partial T}\!&\!=\!&-\frac{1}{2}\nabla_X\!\left[\!\left(1-\frac{\epsilon\gamma_3}{4\pi}r\right)\!r+\epsilon\alpha_1 |\mathbf{w}|^2\!\right]\! \nonumber\\
&\!+\!& \epsilon\alpha_2\mathbf{w}(\nabla_X\!\cdot\mathbf{w})  +\epsilon\alpha_3(\mathbf{w}\cdot\nabla_X)\mathbf{w}+\epsilon\delta\nabla^2_X \mathbf{w} \nonumber\\
&\!+&\!\left[\frac{\gamma_3 r}{\pi}+\epsilon\!\left(\eta_2Q_\eta-\frac{r^2}{6\rho_0^2}\!-\!\frac{\mu |\mathbf{w}|^2}{4\pi^2}\right)\!\right]\! \mathbf{w}. \label{eq7b}
\end{eqnarray}\end{subequations}
Eq.~\eqref{eq7a} is a conservation law for the density disturbance $r(\mathbf{X},T)$. The overall density of particles is $\rho_0=N/\ell^2$, which implies the following constraint on a square box of side $L=\epsilon\ell$:
\begin{equation}
\int_{[0,L]^2} r(\mathbf{X},T)\, d\mathbf{X}=0.\label{eq8}
\end{equation}
The coefficients in Eqs.~\eqref{eq7} are calculated asymptotically for large $M$ \cite{BT18}:
\begin{eqnarray}
&&\alpha_1\!\sim\! -\frac{2\pi \mathbb{r}_0^2\cos\frac{\eta_c}{2}}{\sqrt{\pi M}-\cos\frac{\eta_c}{2}}\!\left(1\!+\!\frac{1}{8M}\right)\!,\nonumber\\
&&\alpha_2\!\sim\!\alpha_1\! +\frac{1}{2\rho_0},\quad\alpha_3\!\sim\!\frac{ -\pi \mathbb{r}_0^2\cos\frac{\eta_c}{2}}{\sqrt{\pi M}-\cos\frac{\eta_c}{2}}\!\left(1\!-\!\frac{1}{8M}\right)\!, \nonumber\\
&&\gamma_0\!=\frac{1}{1\!-\!\frac{1}{\sqrt{\pi M}}\cos\frac{\eta_c}{2}},\quad\rho_0\alpha_3\!\sim\!(1-\gamma_0)\left(M\!-\!\frac{1}{8}\right)\!, \nonumber\\
&&\gamma_3\sim\frac{\pi^2\mathbb{r}_0^2}{2M}=\frac{\pi}{2\rho_0},\quad\delta=\frac{1+\frac{1}{\sqrt{\pi M}}\cos\frac{\eta_c}{2}}{8\!\left(1-\frac{1}{\sqrt{\pi M}}\cos\frac{\eta_c}{2}\right)\!},\nonumber\\
&&\frac{\mu}{4\pi^2}\!\sim\!\frac{M}{4\rho_0^2\!\left(1\!-\!\frac{1}{\sqrt{\pi M}}\cos\frac{\eta_c}{2}\right)\!\!}=\frac{M\gamma_0}{4\rho_0^2}.\quad \label{eq9}
\end{eqnarray}

Eqs.~\eqref{eq7} are solved on a box of side $L=\epsilon\ell$ with periodic boundary conditions and with the additional condition Eq.~\eqref{eq8}. Parameters in the model are $\rho_0$, $M$, $\epsilon$, $L=\epsilon\ell$. Fixing $L$,  $\rho_0$ and $\epsilon$, we find the particle number $N=\rho_0\ell^2= \rho_0L^2/\epsilon^2$. Similarly, fixing $M$ yields $\mathbb{r}_0=\sqrt{M/(\rho_0\pi)}$. To compare direct numerical simulations of the VM with the result of numerically integrating the bifurcation equations, we need to use the previous particle number and the box side $\ell=L/\epsilon$. In the $X$-coordinates the radius of influence is $R_0=\epsilon\mathbb{r}_0$.

\subsubsection{Bifurcating uniform stationary solution}
Eqs.~\eqref{eq7} and \eqref{eq8} have the following uniform solution:
\begin{subequations}\label{eq10}
\begin{eqnarray}
&&r_u=0,\quad \mathbf{w}_u= 2\pi\sqrt{\frac{\eta_2Q_\eta}{\mu}}\, (\cos\Upsilon,\sin\Upsilon), \label{eq10a}\\
&&Q_\eta\!=\!\left.\frac{\partial Q_1}{\partial\eta}\right|_{Q_1=1}\sim-\frac{1}{\eta_c}\!\left(1-\frac{\sqrt{\pi M}}{2}\cos\frac{\eta_c}{2}\right)\!, \label{eq10b}
\end{eqnarray}\end{subequations}
where $\Upsilon$ is a real number and $Q_1$ is the multiplier that surpasses 1 as $\eta<\eta_c$ \cite{BT18}. As $Q_\eta<0$ on the critical line and $\mu>0$, we have $\eta=-1$ and the uniform solution of Eqs.~\eqref{eq10} exists for $\eta<\eta_c$ and is linearly stable. For $\eta>\eta_c$, the uniform solution $r=0$, $\mathbf{w}=\mathbf{0}$ is linearly stable and it becomes unstable for $\eta<\eta_c$.

\subsubsection{Hyperbolic system and oscillations}
For $\epsilon=0$, Eqs.~\eqref{eq7} become
\begin{eqnarray}
\frac{\partial r}{\partial T}+\nabla_X\!\cdot\!\mathbf{w} =0,\quad
\frac{\partial \mathbf{w}}{\partial T}=-\frac{1}{2}\nabla_X r +\frac{1}{2\rho_0}\,r \mathbf{w}, \label{eq11}
\end{eqnarray}
where we have used Eq.~\eqref{eq8}. Eqs.~\eqref{eq11} have to be solved on the square box of side $L$ with periodic boundary conditions and the integral constraint Eq.~\eqref{eq8}. 

\paragraph{Oscillation frequencies.} Linearizing Eqs.~\eqref{eq11} about $r=0$ and $\mathbf{w}=\mathbf{w}_0$ (constant  nonzero vector, which may differ from $\mathbf{w}_u$), we obtain
\begin{eqnarray}
\frac{\partial\tilde{r}}{\partial T}+\nabla_X\!\cdot\!\mathbf{\tilde{w}}=0,\quad
\frac{\partial \mathbf{\tilde{w}}}{\partial T}=-\frac{1}{2}\nabla_X\tilde{r}+\frac{1}{2\rho_0}\mathbf{w}_0\tilde{r}. \label{eq12}
\end{eqnarray}
By differentiating the first equation and eliminating $\mathbf{\tilde{w}}$ by means of the second, we find the wave equation:
\begin{equation}
\frac{\partial^2\tilde{r}}{\partial T^2}=\frac{1}{2}\nabla_X^2\tilde{r} -\frac{1}{2\rho_0}\, \mathbf{w}_0\cdot\nabla_X\tilde{r}. \label{eq13}
\end{equation}
The change of variable $\tilde{r} = e^{\mathbf{w}_0\cdot\mathbf{X}/(2\rho_0)} R$ produces the Klein-Gordon equation:
\begin{equation}
\frac{\partial^2R}{\partial T^2}=\frac{1}{2}\nabla_X^2R -\frac{|\mathbf{w}_0|^2}{8\rho_0^2}\, R.\label{eq14}
\end{equation}
For periodic boundary conditions, $R(\mathbf{X},T)=\sum_{n,m}R_{n,m}(T)e^{i\mathbf{K}_{n,m}\cdot\mathbf{X}}$, and the coefficients $R_{n,m}(T)$ solve the equation of a linear oscillator with frequency
\begin{equation}
\omega_{n,m}\!=\sqrt{\frac{1}{2}|\mathbf{K}_{n,m}|^2 \!+\frac{|\mathbf{w}_0|^2}{8\rho_0^2}}, \quad\mathbf{K}_{n,m}\!=\frac{2\pi}{L}(n,m). \label{eq15}
\end{equation}
Note that the frequency $\omega_{n,m}$ mixes frequencies corresponding to the acoustic velocity $1/\sqrt{2}$ [cf. $|\mathbf{w}_0|=0$ in Eq.~\eqref{eq14}] with the fundamental mode of frequency $|\mathbf{w}_0|/(2\sqrt{2}\rho_0)$ corresponding to $n=m=0$. Note that there may be solutions of the nonlinear system \eqref{eq11} that oscillate with these frequencies about an envelope with approximate amplitude $|\mathbf{w}_0|$ that varies on a longer time scale. We will find such solutions in our numerical simulations.

\section{Hyperbolic system in one dimensional geometry}\label{sec:4}
In the limit as $\epsilon\to 0+$, the solutions of the bifurcation equations \eqref{eq7} can be approximated by the hyperbolic system of Eqs.~\eqref{eq11} for times of order $1/\epsilon$. To connect the 2D band formation in the direct simulations of the VM with the bifurcation equations, we first consider the 1D numerical simulations of the hyperbolic system. 

For initial conditions that are independent of $Y$, with $\mathbf{X}=(X,Y)$, $r=r(X,0)$, we have $\mathbf{w}=(w(X,0),0)$, and Eqs.~\eqref{eq11} are 1D
\begin{eqnarray}
r_T+w_X=0,\quad w_T=-\frac{1}{2}r_X +\frac{rw}{2\rho_0}. \label{eq16}
\end{eqnarray}
Here and henceforth subscripts mean partial derivatives with respect to the corresponding variable. We write these equations in terms of Riemann invariants \cite{lax73}, namely $\xi_\pm=(r\pm\sqrt{2}w)/2$, thereby obtaining
\begin{eqnarray}
\left(\begin{array}{c}
\xi_+\\ \xi_-\\ \end{array}\right)_T+ \frac{1}{\sqrt{2}}\left(\begin{array}{c}
\xi_+\\ -\xi_-\\ \end{array}\right)_X= \frac{\xi_+^2-\xi_-^2}{4\rho_0}\left(\begin{array}{c}
1\\ -1\\ \end{array}\right)\!. \label{eq17}
\end{eqnarray}
Eqs.~ \eqref{eq17} with $u = -\xi_+$ and $v = -\xi_-$ are the Carleman system of equations \citep{god71}, but with negative values of $u$ and $v$. 

The Riemann invariants $\xi_\pm$ are waves traveling with velocities $\pm 1/\sqrt{2}$. The linear combinations $r=\xi_++\xi_-$ and $w=(\xi_+-\xi_-)/\sqrt{2}$ are therefore {\em standing waves}.  We use periodic initial conditions for $r(X,0)$ and $w(X,0)=1$. The initial profile $r(X,0)$ has to satisfy the constraint Eq.~\eqref{eq8}, we set $L=\pi$, and 
\begin{eqnarray}
r(X,0)=\!\left\{ \begin{matrix} 
&\hspace{-2.5cm}\sin(\nu X) &\mbox{even }\nu\in\mathbb{Z}, \\
&\sin(\nu X) - \frac{12}{\pi^3 \nu}X(\pi-X) &\mbox{odd}\,\nu\in\mathbb{Z}. 
\end{matrix}\right.\quad\,\,  \label{eq18}
\end{eqnarray}

\begin{figure*}[htp]
\begin{center}
\includegraphics[width=\textwidth]{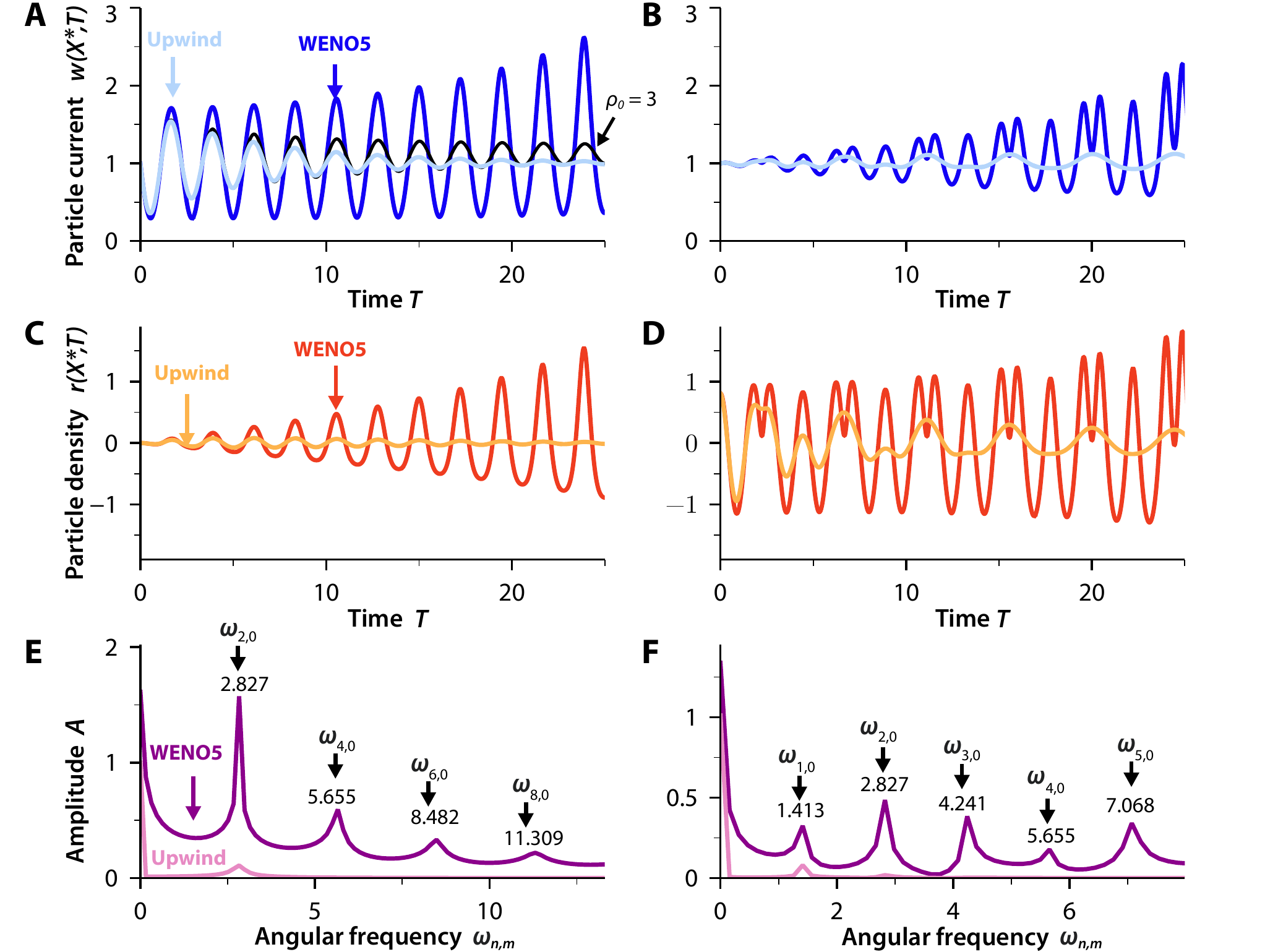}
\end{center}
\caption{Simulation of the 1D hyperbolic system~(\ref{eq16}) for even wavenumber $\nu=4$ (left column) and odd wavenumber $\nu=5$ (right column), $\rho_0=10$ and $L=\pi$, showing the time evolution of the density and disturbance density of the bifurcation equations. \textbf{(A)} Current $w(X^*,T)$ for $X^*=L/2$ as a function of time $T$ for upwind scheme and WENO5 (darker line). \textbf{(B)} Same as \textbf{(A)} but with $\nu = 5$ in the initial condition. \textbf{(C)} density $r(X^*,T)$ as a function of time $T$ for upwind scheme (dotted line) and WENO5 (solid line). \textbf{(D)} Same as \textbf{(C)} but with $\nu = 5$ in the initial condition. \textbf{(E)}-\textbf{(F)} Amplitude of the Fourier spectrum of $w(X^*,T)$ vs. frequency for upwind scheme and WENO5.
\label{fig1}}
\end{figure*}
To numerically solve Eqs.~\eqref{eq17}-\eqref{eq18} and study the behavior of their solutions, we have used a first order (in time and space) upwind method. The explicit upwind scheme requires a Courant-Friedrichs-Levy (CFL) condition on the time step, $\Delta t$, in terms of the wave speed $c=1/\sqrt{2}$ and the space grid size $\Delta x$, \cite{lax73}
\begin{eqnarray}
c \frac{\Delta t}{\Delta x}\leq \kappa,\quad 0<\kappa<1. \label{eq19}
\end{eqnarray}
Using the upwind finite difference scheme for the spatial terms and the Euler explicit algorithm to evolve in time, upwind lines in Figs.~\ref{fig1}\textbf{\textbf{(A)}}-\ref{fig1}\textbf{(D)} show a strong dissipation due to numerical noise and the oscillations over time disappear. Indeed, for initial conditions $r(X,0)$ with even and odd number of maxima, the current density and the density disturbance are both damped by the numerical noise of the upwind scheme: As $T\to\infty$, $r(X^*,T)$ and $w(X^*,T)$ go to 0 and 1 (the spatial averages of their initial conditions), respectively. Moreover, as we can see in Fig.~\ref{fig1}\textbf{(A)}, decreasing the density $\rho_0$ in the system (from 10 to 3), has a non-stabilizing effect in the amplitude of the standing waves (See black line in Fig.~\ref{fig1}\textbf{(A)} for $\rho_0=3$). 

If we want to capture correctly these wave patterns, we need to apply high accuracy numerical methods. For this purpose, we have also formulated a high order scheme based on the fifth-order weighted essentially non oscillatory reconstruction procedure (WENO5) for the spatial variables and a third order Runge-Kutta method for time integration (See Section~\ref{section_vi} for details about Upwind and WENO5 schemes). The WENO5 procedure was designed to get fifth order accuracy in space using a nonlinear convex combination of essentially nonoscillatory parabolas (see Ref.~\cite{jia96} for details), whereas the third order Runge-Kutta (RK3) method allows the maximum time stepsize dictated by the first-order upwind scheme \cite{shu89}. The numerical results obtained by WENO5 are shown in Fig.~\ref{fig1} with darker lines. We compute the numerical approximation for $N$=100, $\Delta x$ = 0.0314 and $\Delta t=10^{-4} $. 
This high order numerical method eliminates the numerical dissipation of the first order upwind scheme, as illustrated by the comparison between them shown in Fig.~\ref{fig1}. Moreover, the higher order scheme reveals that the amplitudes of the current density and the density disturbance both increase with time as the traveling waves comprising them pass through $X^*=L/2$. 
For even $\nu$, Figs.~\ref{fig1}\textbf{(A)} and \ref{fig1}\textbf{(C)} show that the wave keeps its shape and the minima and maxima are better captured whereas the number of maxima change, as depicted in Figs.~\ref{fig1}\textbf{(B)} and \ref{fig1}\textbf{(D)} for odd $\nu$. The variable number of peaks yields additional peaks in the Fourier spectrum, as shown by a comparison of Figs.~\ref{fig1}\textbf{(E)} and \ref{fig1}\textbf{(F)}. For $\rho_0=10$, $L=\pi$, Eq.~\eqref{eq15} predicts peaks with frequencies 1.4147, 2.8286, 4.2428, 5.6570, and 7.0712, corresponding to $m=0$ and $n=1,2,3,4,5$, respectively. These are close to the numerically obtained values 1.413, 2.827, 4.241, 5.655, and 7.068 appearing in Fig.~\ref{fig1}\textbf{(F)} for $\nu=5$. For an initial condition with even $\nu=4$, only the peaks corresponding to even $n=2,4,6,8$ appear in Fig.~\ref{fig1}\textbf{(E)}, at the same frequencies as in Fig.~\ref{fig1}\textbf{(F)}.

\section{Parabolic system}\label{sec:5}
In this section, we compare numerical solutions of the full 1D model in Eqs.~\eqref{eq7} for different values of $\epsilon$ to numerical solutions of the hyperbolic approximation Eq.~\eqref{eq11}. For 1D geometry, we write Eqs.~\eqref{eq7} in terms of the Riemann invariants and integrate them using a WENO5 method. We compare these numerical solutions to those obtained by integrating the full 1D model Eqs.~\eqref{eq7} using the Fourier spectral method and a central difference scheme. We also solve numerically Eqs.~\eqref{eq7} for a 2D geometry and show that the model describes periodic patterns of traveling densities caused by moving active particles. 

\subsection{One dimensional geometry}
The one-dimensional version of Eqs.~\eqref{eq7} is
\begin{subequations}\label{eq20}
\begin{eqnarray}
&&r_T+w_X+\frac{\epsilon}{4\rho_0}(rw)_X=0,\label{eq20a}\\
&&w_T+\frac{1}{2}r_X -\frac{\epsilon}{8\rho_0}r r_X-\epsilon\!\left(\frac{1}{2\rho_0} +\alpha_3\!\right)\! ww_X\nonumber\\
&&\quad=\frac{rw}{2\rho_0}+\epsilon\delta w_{XX} +\epsilon\!\left(\Gamma(r)-\frac{\mu w^2}{4\pi^2}\right)\! w,\quad \label{eq20b}\\
&&\Gamma(r)=\eta_2Q_\eta-\frac{r^2}{6\rho_0^2},\label{eq20c}
\end{eqnarray}
\end{subequations}
in which we have used Eq.~\eqref{eq9}. In order to compare the numerical solutions of the Carleman system \eqref{eq17} with Eqs.~\eqref{eq20}, we need to write the latter in terms of the Riemann invariants $\xi_\pm$. In vector form, Eqs.~\eqref{eq20} become
\begin{eqnarray}
&&\!\left(\begin{array}{c} r\\w\\ \end{array}\right)_T+\left(\begin{array}{cc} \frac{\epsilon w}{4\rho_0}& 1+\frac{\epsilon r}{4\rho_0}\\ \frac{1}{2}-\frac{\epsilon r}{8\rho_0}& -\epsilon\!\left(\frac{1}{2\rho_0}+\alpha_3\!\right)\!w\end{array}\right)\!\left(\begin{array}{c} r\\w\\ \end{array}\right)_X \nonumber\\
&&\quad\quad =\left(\begin{array}{c} 0\\ \frac{rw}{2\rho_0}+\epsilon\delta w_{XX} +\epsilon\!\left(\Gamma(r)-\frac{\mu w^2}{4\pi^2}\right)\! w\\ \end{array}\right)\!.\quad
\label{eq21}
\end{eqnarray}
The convection matrix on the left hand side has eigenvalues and left eigenvectors:
\begin{eqnarray}
&&\lambda_\pm \!= -\frac{\epsilon(1+4\rho_0\alpha_3)w}{8\rho_0}\pm\sqrt{\frac{1}{2}+\epsilon^2 \frac{(3+4\rho_0\alpha_3)^2w^2\!-\!2r^2}{64\rho_0^2} },\nonumber \\
&&\mathbf{l}_\pm= \left(\begin{array}{c} \frac{1}{2}-\frac{\epsilon r}{8\rho_0}\\ \lambda_\pm-\frac{\epsilon w}{4\rho_0}\\ \end{array} \right)\!.   \label{eq22}
\end{eqnarray}
For the system of Eq.~\eqref{eq21} without right hand side, the Riemann invariants are parallel to the left eigenvectors  $\mathbf{l}_\pm$ \cite{lax73}, therefore they satisfy
\begin{eqnarray}
\frac{\xi_{\pm,r}}{\frac{1}{2}-\frac{\epsilon r}{8\rho_0}}=\frac{\xi_{\pm,w}}{\lambda_\pm-\frac{\epsilon w}{4\rho_0}}= \chi(r,w). \label{eq23}
\end{eqnarray}
Here $\chi(r,w)$ is an integrating factor such that
\begin{eqnarray}
d\xi_\pm=\left(\frac{1}{2}-\frac{\epsilon r}{8\rho_0}\right)\chi\, dr+\left(\lambda_\pm-\frac{\epsilon w}{4\rho_0}\right) \chi\, dw. \label{eq24}
\end{eqnarray}
Integrating $d\xi_\pm=0$ implies solving the equations:
\begin{eqnarray}
\frac{dr}{dw}=\frac{2}{1-\frac{\epsilon r}{4\rho_0}}\left(\frac{\epsilon w}{4\rho_0}-\lambda_\pm\right)\!, \label{eq25}
\end{eqnarray}
and the resulting functions $\xi_\pm(r,w;\epsilon)=M_\pm$ (constant) are the Riemann invariants. We find approximate solutions by expanding the right hand side of Eq.~\eqref{eq25} in powers of $\epsilon$. We obtain: 
\begin{eqnarray}
\xi_\pm =\frac{1}{2}\left[r\pm \sqrt{2}w- \frac{\epsilon}{8\rho_0}r^2-\epsilon\left(\frac{3}{8\rho_0}+\frac{\alpha_3}{2}\right)w^2\right]\!,\quad \label{eq26}
\end{eqnarray}
up to $O(\epsilon)$ terms. As Eq.~\eqref{eq21} has source terms, the functions $\xi_\pm$ are no longer invariant, although we will still call them Riemann invariants. After some algebra, we can obtain equations for $\xi_\pm$ by inserting Eq.~\eqref{eq26} into Ea.~\eqref{eq21} and ignoring $O(\epsilon^2)$ terms:
\begin{widetext}
\begin{subequations}\label{eq27}
\begin{eqnarray}
&&\!\left(\begin{array}{c} \xi_+\\ \xi_-\\ \end{array}\right)_T+\frac{1}{\sqrt{2}}\!\left[\bm{\sigma}_z-\epsilon\left( \frac{1}{8\rho_0}+\frac{ \alpha_3}{2}\right)\!(\xi_+-\xi_-)\mathbb{I}\right]\!\left(\begin{array}{c} \xi_+\\ \xi_-\\ \end{array}\right)_X\! 
=\!\left[\frac{\delta\epsilon}{2}(\xi_+-\xi_-)_{XX}+S(\xi_+,\xi_-)\right]\!\left(\begin{array}{c} 1\\ -1\\ \end{array}\right)\nonumber\\
&&\quad\quad\quad\quad +\,\frac{\epsilon(\xi_+\!-\xi_-)}{4\rho_0} 
\left[ \frac{1}{2\rho_0}\left(\begin{array}{c}\xi_-^2\\ -\xi_+^2\\ \end{array}\right)\!-\frac{1+4\rho_0\alpha_3}{16\rho_0} (\xi_+-\xi_-)\!\left(\begin{array}{c} \xi_++3\xi_-\\ \xi_-+3\xi_+\\ \end{array}\right)\right]\!, \label{eq27a}\\
&& S(\xi_+,\xi_-)= \frac{\xi_+\!^2-\xi_-^2}{4\rho_0} + \frac{\epsilon (\xi_+-\xi_-)}{2}\!\left[\Gamma(\xi_+ +\xi_-) -\mu\,\frac{(\xi_+-\xi_-)^2}{8\pi^2}\right]\!\!, \quad
\bm{\sigma}_z=\left(\begin{array}{cc} 1&0\\ 0&-1\\ \end{array}\right)\!, \quad\mathbb{I}=\left(\begin{array}{cc}1& 0\\ 0& 1\\\end{array}\right)\!. \label{eq27b}
\end{eqnarray}
\end{subequations}
\end{widetext}
\begin{figure*}[htp]
\begin{center}
\includegraphics[width=\textwidth]{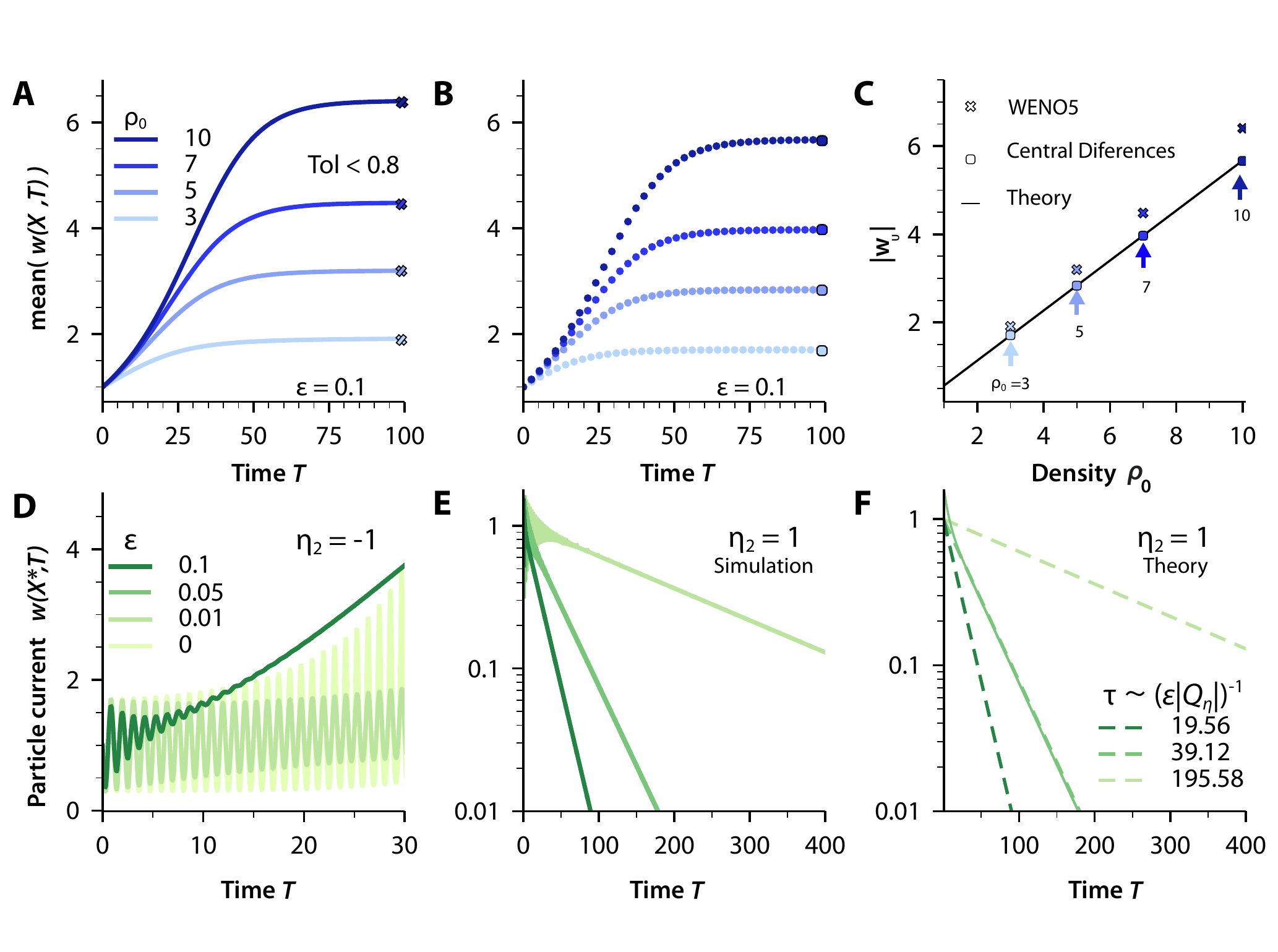}
\end{center}
\vspace*{-1cm}
\caption{Numerical simulation of the parabolic system of equations with $L=\pi$, $M=7$, $\rho_0$ and $\epsilon$ as indicated in the panels, and initial condition $r = \sin(\nu X)$, $\nu = 8$, $w=1$. Space averages $\langle w\rangle$ as functions of time for the values of $\rho_0=3,5,7,10$, $\epsilon=0.1$, $\eta_2=-1$ calculated by \textbf{(A)} numerical integration of Riemann invariants Eq.~\eqref{eq27}; \textbf{(B)} direct integration of the parabolic system Eqs.~\eqref{eq20} using centered differences. \textbf{(C)} Comparison of the numerically calculated $|\mathbf{w}_u|$ by the previous methods and by Eq.~\eqref{eq10} (theory). \textbf{(D)} $w(X^*,T)$ calculated using the (WENO5, RK3) method for $\eta_2= - 1$, $\rho_0=10$, and $\epsilon=0, 0.01, 0.1$. \textbf{(E)} Same as \textbf{(D)} for $\eta_2=1$ and $\epsilon=0.01, 0.05, 0.1$. \textbf{(F)} Calculation of the relaxation time for the upper envelope of the oscillations in Panel \textbf{(E)}; the theoretical value is $(\epsilon |Q_\eta|)^{-1}\approx 1.9558/\epsilon$. 
\label{fig2}}
\end{figure*}

We solve Eqs.~\eqref{eq27} using a WENO5 for the spatial variables and a third order Runge-Kutta method for time integration. Additionally,  in order to compare the numerical solutions of Eqs.~\eqref{eq27} (WENO5,RK3), we have also used a Fast Fourier spectral method and a central difference scheme to solve the full Eqs.~\eqref{eq11} without approximations. The numerical results of the density and disturbance density are displayed in Fig.~\ref{fig2} for WENO5 scheme (solid line), central difference scheme (dotted line), and Fourier spectral method (that has the same results as the central difference scheme).

Figs.~\ref{fig2}\textbf{(A)} and \ref{fig2}\textbf{(B)} compare the numerical solution of Eqs.~\eqref{eq27} (WENO5, RK3) with a direct numerical solution of Eqs.~\ref{eq20} for different values of the density $\rho_0$ and $\epsilon=0.1$. Fig.~\ref{fig2}\textbf{(C)} displays the module of the stationary solution $|\mathbf{w}_u|$ once it has reach the plateau in Figs.~\ref{fig2}\textbf{(A)}-\textbf{(B)}. The black curve comes from theory whereas the crosses and squares come from (WENO5,RK3) and central difference schemes respectively. Differences between numerical schemes are due to the approximate Riemann invariants, and increase with $\rho_0$. 

Figs.~\ref{fig2}\textbf{(D)} and \ref{fig2}\textbf{(E)} show $w(L/2,T)$ vs time for $\eta_2=-1$ and $\eta_2=1$, respectively, for $\epsilon=0, 0.01, 0.05, 0.1$ ($\epsilon$ is the distance to the critical bifurcation point). Fig.~\ref{fig2}\textbf{(D)} shows that the greater $\epsilon$ is, the faster the oscillations disappear and the solution approaches its stationary value more rapidly. Fig.~\ref{fig2}\textbf{(F)} compares the relaxation times obtained from the space uniform version of Eq.~\eqref{eq7b} (with $r=0$) and Eq.~\eqref{eq10b}  with those obtained from the numerical solutions. The agreement between the theoretical and numerical relaxation times is excellent.

\begin{figure*}[htp]
\begin{center}
\includegraphics[width=\textwidth]{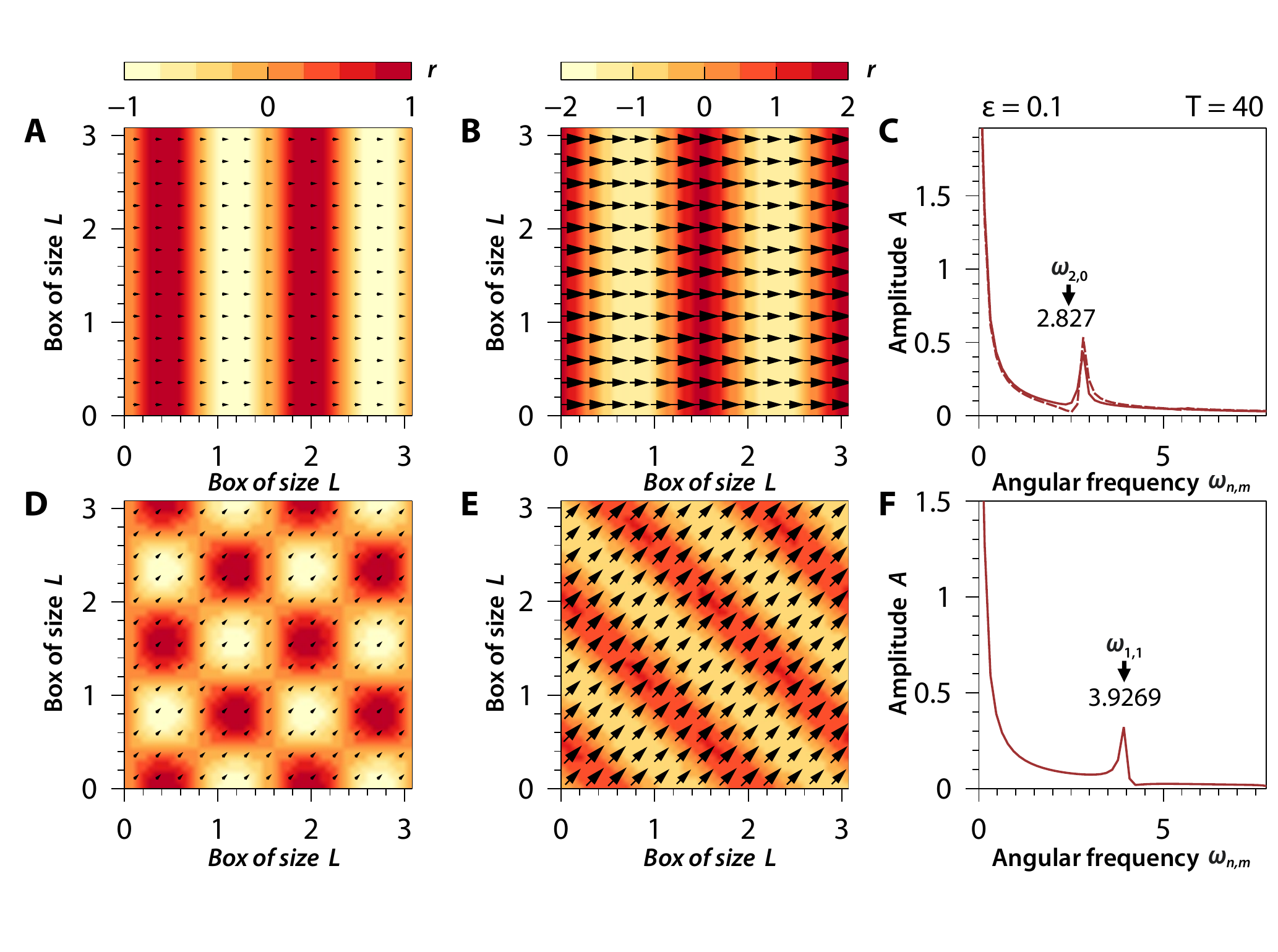}
\end{center}
\vspace*{-1cm}
\caption{Numerical simulations of the parabolic system~(\ref{eq7}) for wavenumber $\nu=4$, $\rho_0=10$, $\epsilon=0.1$, and $L=\pi$. \textbf{(A)} Density plot of the initial condition $\mathbf{(I)}$; the arrows indicate the direction of $\mathbf{w}$, \textbf{(B)} Same as \textbf{(A)} for $T=40$. \textbf{(C)} Amplitude of the Fourier spectrum of $w(X^*,Y^*,T)$ vs. frequency showing a peak at the frequency 2.827 [corresponding to $n=2$, $m=0$ in Eq.~\eqref{eq15}] for a 2D spectral method (solid black line) and 1D WENO5 (dotted red line). \textbf{(D)} Same as \textbf{(A)} for the initial condition $\mathbf{(II)}$. \textbf{(E)} Same as \textbf{(D)} for $T=40$. \textbf{(F)} Same as \textbf{(C)} for the initial condition $\mathbf{(II)}$ of \textbf{(D)}; the peak frequency is 3.9269 corresponding to $n=m=1$ in Eq.~\eqref{eq15}.
\label{fig3}}
\end{figure*}
\begin{figure*}[htp]
\begin{center}
\includegraphics[width=\textwidth]{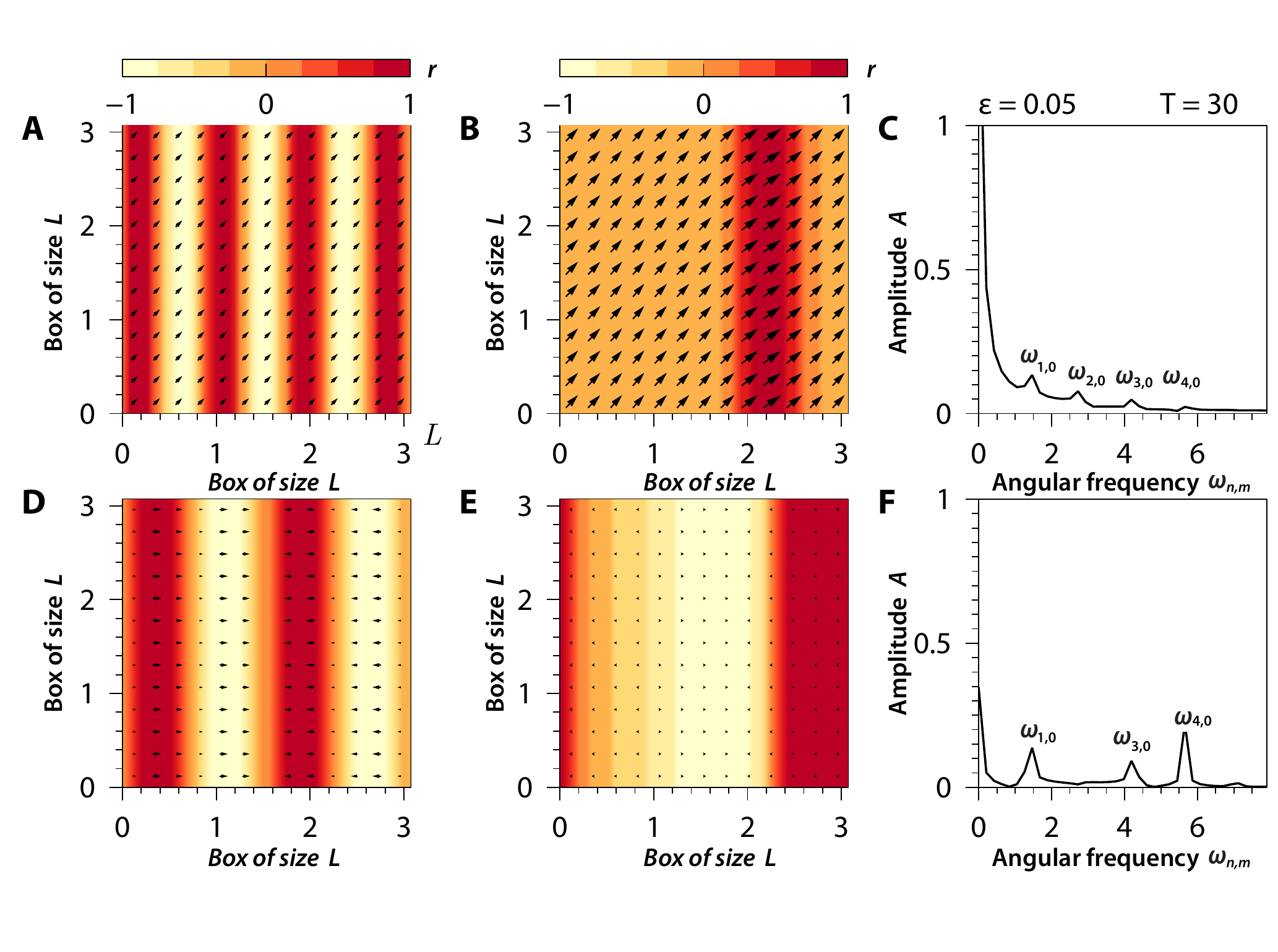}
\end{center}
\vspace*{-1cm}
\caption{Numerical simulations of the parabolic system~(\ref{eq7}) for wavenumber $\nu=7$, $\rho_0=10$, $\epsilon=0.05$, and $L=\pi$. \textbf{(A)} Density plot of the initial condition $\mathbf{(III)}$; the arrows indicate the direction of $\mathbf{w}$, \textbf{(B)} Same as \textbf{(A)} for $T=40$, \textbf{(C)} Amplitude of the Fourier spectrum of $\mathbf{w}(X^*,Y^*,T)$ vs. frequency showing different peaks using the 2D spectral method. \textbf{(D)} Same as \textbf{(A)} for the initial condition $\mathbf{(IV)}$ with $\nu=4$. \textbf{(E)} Same as \textbf{(B)} for the initial condition in \textbf{(D)}. \textbf{(F)} Same as \textbf{(C)} for the initial condition in \textbf{(D)}.
\label{fig4}}
\end{figure*}

\subsection{Two dimensional geometry}
Here, we describe band formation using the two-dimensional parabolic equations presented in Eqs.~\eqref{eq7}.
To further characterize the dynamics, we consider a 2D geometry on a square domain $[0,\pi]\times[0,\pi]$. In particular, we examine the average density and velocity of the system considering the following four initial conditions that mimic the behavior found in the direct simulations of the model described in Section~\ref{sec:2}:
\begin{subequations}\label{eq28}
\begin{eqnarray}
\hspace{-1.3cm}\mathbf{(I)}:=\!\left\{ \begin{matrix} 
r(X,Y, 0)= &\hspace{-0.2cm}\sin(\nu X)&\mbox{even }\nu\in\mathbb{Z}, \\
\mathbf{w}(X,Y, 0)=&\hspace{-0.6cm}(1,0) &. 
\end{matrix}\right.\quad\,\,  \label{eq28a}
\end{eqnarray}
\begin{eqnarray}
\mathbf{(II)}:=\!\left\{ \begin{matrix} 
r(X,Y, 0)= &\hspace{-0.2cm}\cos(\nu Y)\sin(\nu X)&\mbox{even }\nu\in\mathbb{Z}, \quad\\
\mathbf{w}(X,Y, 0)=&\hspace{-0.9cm}(1,1)/\sqrt(2) &. 
\end{matrix}\right.\quad\,\,  \label{eq28b}
\end{eqnarray}
\begin{eqnarray}
\hspace{-1.3cm}\mathbf{(III)}:=\!\left\{ \begin{matrix} 
&\hspace{-2.15cm}\sin(\nu X) - \frac{12}{\pi^3 \nu}X(\pi-X) &\mbox{odd }\nu\in\mathbb{Z}, \\
\mathbf{w}(X,Y, 0)=&\hspace{-0.6cm}(1,0) &. 
\end{matrix}\right.\quad\,\,  \label{eq28c}
\end{eqnarray}
\begin{eqnarray}
\mathbf{(IV)}:=\!\left\{ \begin{matrix} 
r(X,Y, 0)= &\hspace{-3cm}\sin(\nu X)&\hspace{-3cm}\mbox{even }\nu\in\mathbb{Z}, \\
\mathbf{w}(X,Y, 0)=&\hspace{-0.1cm}(r(X,Y,0)^2 \mbox{sign}(\frac{\pi}{2}-X),\,0)&\hspace{-0.2cm}. 
\end{matrix}\right.\quad\,\,  \label{eq28d}
\end{eqnarray}
\end{subequations}
We integrate numerically Eqs.~\eqref{eq7} using the Fourier Spectral method in space and the Euler explicit algorithm for the time evolution. Figs.~\ref{fig3} displays the resulting density contour profile. The arrows represent the vector field of the current density, $w(x,t)$. We observe that the solution is independent of $Y$ for the initial condition $\mathbf{(I)}$, and Figs.~\ref{fig3}\textbf{(A)}-\ref{fig3}\textbf{(C)} produce the same result as in the 1D case. For the central point of the square, Fig.~\ref{fig3}\textbf{(C)} shows that, at time $T=40$, the 2D spectral method (dashed line) yields the same amplitude of the Fourier spectrum as the 1D solid line (WENO5,RK3) method. The peak frequency at 2.827 coincides with the numerical value given by setting $n=2$, $m=0$ in Eq.~\eqref{eq15}. In addition, we observe that the Fourier spectrum in Fig.~\ref{fig3}\textbf{(C)} has less peaks than previously calculated for the hyperbolic system and depicted in Fig.~\ref{fig1}\textbf{(E)}. The reason is that the value of $\epsilon$ is sufficiently large for the oscillations to end at a relatively short time, cf. Fig.~\ref{fig2}. Thus, the Fourier spectrum does not collect all the frequencies appearing in the hyperbolic case. For initial condition $\mathbf{(II)}$, Figs.~\ref{fig3}\textbf{(D)}-\ref{fig3}\textbf{(F)} exhibit similar waves traveling along the direction $Y=X$ with the same velocity. The Fourier spectrum in Fig.~\ref{fig3}\textbf{(F)} yields a peak with frequency 3.929 given by setting $m=1$ and $n=1$ in Eq.~\eqref{eq15}. Of course, the initial condition $\mathbf{(II)}$ produces 2D patterns without correspondence with the 1D case of initial condition $\mathbf{(I)}$.

We have also analyzed the behavior of the bands for odd wavenumber. As in the 1D case, their evolution does not preserve the number of maxima and minima, unlike what happens in the  case of even wavenumber. For 2D geometry and initial condition $\mathbf{(III)}$, Figs.~\ref{fig4}\textbf{(A)}-\textbf{(C)} show how the waves migrate from left to right and finally merge into a single larger wave. For the initial condition $\mathbf{(IV)}$, we have two waves moving in opposite directions, as shown by the arrows indicating the direction of $w_x(X,Y,T)$ in Fig.~\ref{fig4}\textbf{(D)}. After a sufficiently large time, left and right moving waves merge into a single band; see Fig.~\ref{fig4}\textbf{(E)}. 

\begin{figure*}[htp]
\begin{center}
\includegraphics[width=\textwidth]{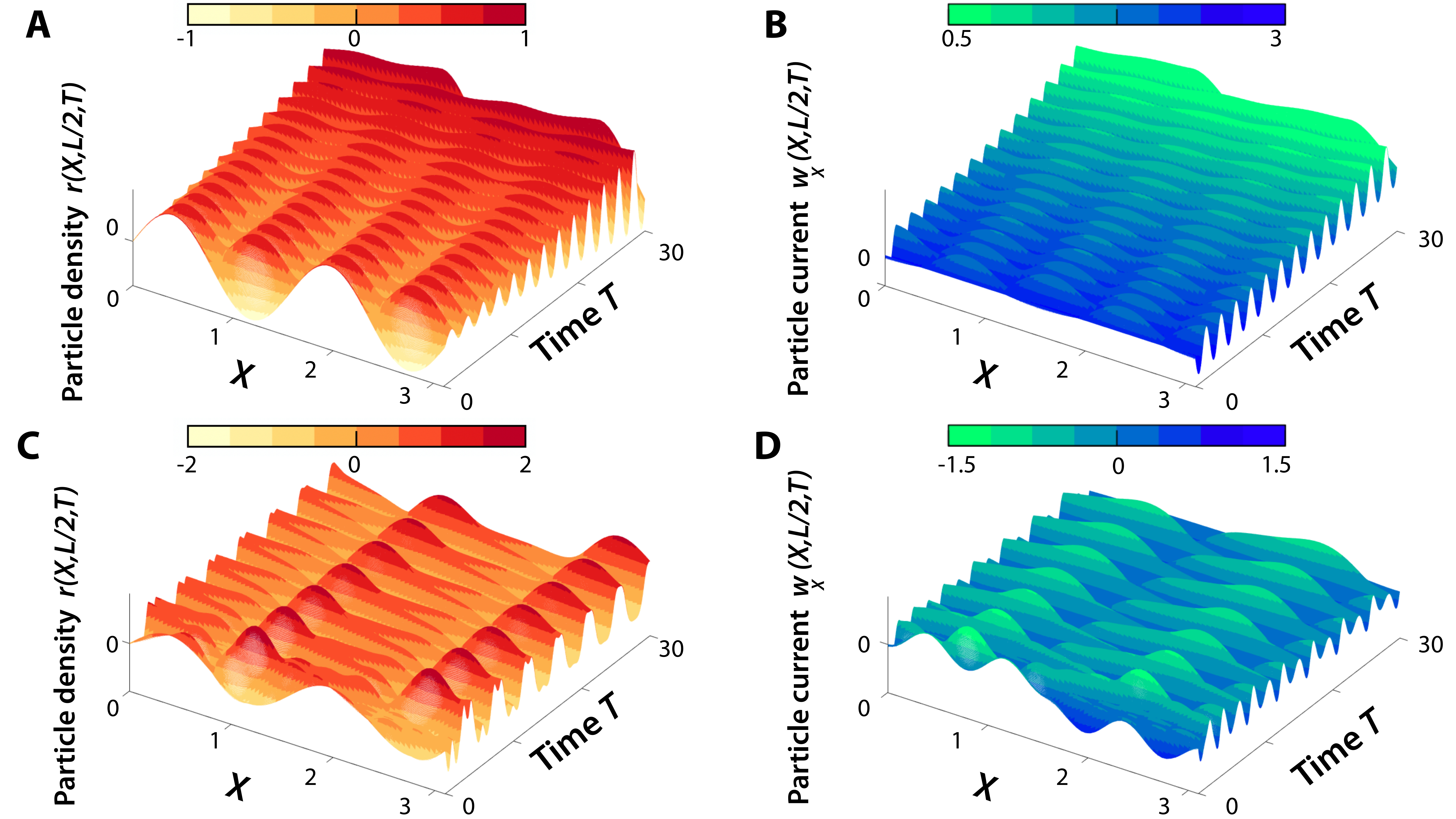}
\end{center}
\vspace{-0.7cm} 
\caption{Numerical solutions of the parabolic system \eqref{eq7} for $\nu=4$ and other parameters as in Fig.~\ref{fig4}. \textbf{(A)} Surface plots of the density disturbance $r(X,Y=L/2,T)$ and \textbf{(B)} for the current density $w_x(X,L/2,T)$ on the $z$ direction for the initial condition $\mathbf{(I)}$. These functions are the same for any value of $Y$. (c-d) Same as (a-b) for the initial condition $\mathbf{(IV)}$. 
\label{fig5}}
\end{figure*}

The initial condition $\mathbf{(IV)}$ produces waves moving in opposite directions. To better understand this behavior, we represent the average density and current density in a waterfall plot for different times; cf Fig.~\ref{fig5}. For the initial condition $\mathbf{(I)}$,  the numerical solution displays waves moving from left to right as shown in Figs.~\ref{fig5}(A-B). Figs.~\ref{fig5}(C-D) shows that the initial condition $\mathbf{(IV)}$ yields waves moving from left to right and waves moving from right to left. These waves collide at the middle of the computational domain and are regenerated some time later by the periodic boundary conditions. As time evolves only two waves survive from the initial four and the overall effect is having an oscillating single band.

\begin{figure*}[htp]
\begin{center}
\includegraphics[width=\textwidth]{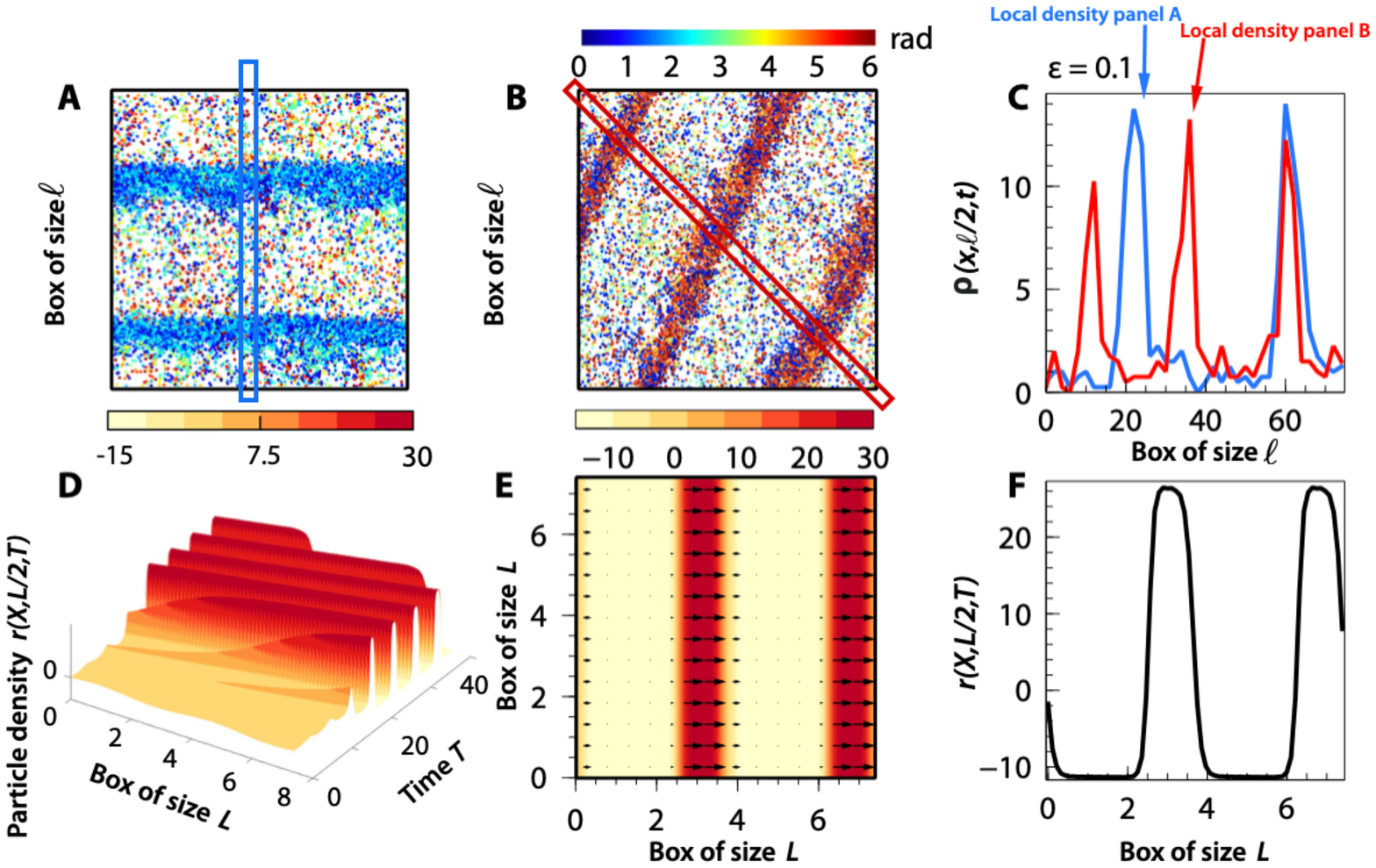}
\end{center}
\vspace{-5.0cm} 
\caption{{\bf(A)}, {\bf(B)} and {\bf (C)} Numerical simulations of the Vicsek model of Eqs.~\eqref{eq1}-\eqref{eq2} for $\ell=74.8998$, $\rho_0=3$, $N=16830$, $\mathbb{r}_0=0.5$, $M=2.3562$ and $\eta=2.6233$ corresponding to parameters $\epsilon=0.1$, $L=\ell\epsilon=7.48998$, $R_0=\mathbb{r}_0=0.05$ in {\bf (D)}, {\bf (E)} and {\bf (F)} for Eqs.~\eqref{eq7}-\eqref{eq9}. The initial conditions for Eqs.~\eqref{eq1}-\eqref{eq2} are random in space and velocity, and for Eqs.~\eqref{eq7}-\eqref{eq9}  we have used initial condition {\bf (II)} and wavenumber $\nu=4$. We observe the formation of bands with orientations $\pi/2$ and $7\pi/4$ in the numerical simulations of the VM, which are similar to the bands obtained by integrating Eqs.~\eqref{eq1}-\eqref{eq2}. 
\label{fig6}}
\end{figure*}

\section{Comparison with direct Vicsek model simulations}\label{sec:6}
Direct simulations of the Vicsek model described by Eqs.~\eqref{eq1}-\eqref{eq2} produce patterns similar to those predicted by solving the PDEs \eqref{eq7}-\eqref{eq9}. To directly connect the VM simulations to the solutions of the bifurcation equations, we solve both on a two-dimensional square domain. In the VM simulations, the continuum variables $r(X,Y,t)$ and $w(X,Y,T)$ correspond to the local density and the local velocity, respectively. Fig.~\ref{fig6} compares band formation. For direct simulations of the VM, we have considered initial random conditions for the position and velocity of the particles, box size $\ell = 74.8998$, density $\rho_0 = 3$, particle number $N =16830$, radius of interaction $\mathbb{r}_0 = 0.5$, distance to the bifurcating point $\epsilon = 0.1$ and average number of neighbors $M = 2.3562$.  At early times, direct numerical simulations of the Vicsek model start with random initial conditions in space and velocity. Eventually, as particles start migrating, bands with different spatial orientations arise as shown in Figs.~\ref{fig6}(A-B). To further characterize the formation of bands, we obtain the local density of the system along the perpendicular direction to the bands at a fixed time. To this end, we define a rectangle domain of dimensions $2\times74.8998$ and we compute the mean value of the local density within it. As a result, at sufficiently long time we observe that the density of the system is not uniform in space, as shown in Fig.~\ref{fig6}(C). Fig.~\ref{fig6}(E) shows that the same pattern formation appears when solving numerically the 2D parabolic bifurcation equations with parameter values $L = \ell \epsilon = 7.48998$, $R_0 = \mathbb{r}_0 \epsilon =  0.05$, $M =2.3562$, $N =16830$, $\rho_0 =  3$ and $\eta_c = 2.6333$. The simulations also confirm the time evolution of the local density, as shown in Fig.~\ref{fig6}(D). At the initial stage, the system has uniform local density but, after a sufficiently long time, there emerge patterns similar to those found in the agent-based numerical simulations of the stochastic VM.  To understand the accumulation of particles in the planar waves, we compare quantitatively the average  local particle density obtained from direct VM simulations, Fig.~\ref{fig6}(C), to that obtained from numerical solution of the parabolic bifurcation equations, Fig.~\ref{fig6}(F). In both cases, the  particle density is nonuniform and reaches large values along the bands. 
In Fig.~\ref{fig6}(C), the mean value of the amplitude is $11.5$ which is comparable with the amplitude $40\times \epsilon=4$ obtained from Fig.~\ref{fig6}(F).

\section{Conclusions}\label{sec:7}
The 2D Vicsek model describes flock formation presenting a transition between a disorder state to an ordered state at which animal flocks or swarms are formed. Using a molecular chaos assumption, Ihle has derived a discrete time Enskog kinetic equation \cite{ihl11,ihl16} that can be further analyzed by bifurcation theory near the order-disorder transition \cite{bon19}. The amplitude equations obtained from bifurcation theory are a coupled conservation law for the density disturbance and a parabolic equation for the current density. The small parameter $\epsilon$  measures the distance to the bifurcation point, defining the transition from hyperbolic equations ($\epsilon=0$) to parabolic equations ($\epsilon\neq 0$).
Although the ordered phase with band formation has been widely studied in the VM, the comparison between numerical simulations of the stochastic VM and the bifurcation equations remains unexplored. Thus, we have solved the bifurcation equations in 1D and 2D expressing them in terms of the Riemann invariants and proposing different numerical schemes to solve them. \\
We have analyzed numerically the 1D hyperbolic system by using a first order numerical scheme after writing it for the Riemann invariants, which ensure an accurate and correct propagation of wave train solutions. These solutions are damped due to numerical dissipation. Thus, we have used a higher order accurate WENO5 reconstruction procedure in space and a third order accurate Runge-Kutta scheme in time, which mitigate numerical dissipation and permit resolving well the local extrema of the waves. Moreover, numerical simulations of the bifurcation equations using a central difference scheme in space and a Fourier spectral method enable us to test that the results obtained with the WENO5 are accurate. Numerical solutions show that the amplitude of the wave trains increases with time. For $\epsilon\neq 0$, the parabolic terms of the PDEs eventually act and stabilize the solutions. Solving the parabolic system by using finite differences or a spectral scheme produce similar results. Furthermore, numerical simulations of the hyperbolic and parabolic equations enable us to confirm the angular frequencies and the relaxation times previously obtained from linearized equations about a spatially uniform solution.  We have also analyzed numerically the 2D parabolic system by using a Fourier spectral method which also predicts the band formation observed in the direct simulations of the VM. Finally, comparisons between simulations of the agent-based VM and the bifurcation equations show the agreement in the average of local particle density between both frameworks.

\acknowledgements
This work has been supported by the FEDER/Ministerio de Ciencia, Innovaci\'on y Universidades -- Agencia Estatal de Investigaci\'on grants PID2020-112796RB-C22 and PID2020-118236GB-I00, by the Madrid Government (Comunidad de Madrid-Spain) under the Multiannual Agreement with UC3M in the line of Excellence of University Professors (EPUC3M23), and in the context of the V PRICIT (Regional Programme of Research and Technological Innovation).

\appendix
\section{One dimensional geometry}\label{section_vi}
\subsection{Riemann invariants and numerical methods}
\paragraph{ First order numerical scheme} 
We adopt a forward difference scheme in time and a first-order upwind scheme in space to solve the hyperbolic conservation laws of the type 
\begin{equation}\label{a1}
{u}_t + cf(u)_x =  0 
\end{equation}
together with the initial data 
\begin{equation}\label{a2}
u(x,0) = u_0(x).
\end{equation}
This numerical scheme can be also used for equations with forcing terms $g(u, x,t)$ on the right hand side of Eq.~\eqref{a1}. 
Let $u_j^n = u_h(x_j, t_n)$ a numerical approximation of the exact solution $u(x_j,t_n)$, Eqs.~\eqref{a1}-\eqref{a2} in conservation form yield
\begin{equation}
u_j^{n+1} = u^n_j - c\frac{\Delta t}{h}(\hat{f}_{j+1/2}-\hat{f}_{j-1/2}),  \label{a2_b}
\end{equation} 
where $c$ is the wave velocity, $\Delta t$ is the timestep and $h$ is the length between mesh elements. The stability condition for this scheme is the CFL condition 
\[c \frac{\Delta t}{h}\leq \kappa,\quad 0<\kappa<1. \]
The interface fluxes $\hat{f}_{j+1/2} = \hat{f}(u^{n}_{j-k+1},...,u^{n}_{j+k})$ and $\hat{f}_{j-1/2}$ are determined by the direction of the wave propagation. We compute the numerical fluxes in the following way: If the characteristic speed $c$ in our system is positive we use backward differences in space, ($\hat{f}_{j+1/2}= \hat{f}_{j}$ and $\hat{f}_{j-1/2} = \hat{f}_{j-1}$) otherwise, forward differences are used ($\hat{f}_{j+1/2}= \hat{f}_{j+1}$ and $\hat{f}_{j-1/2}= \hat{f}_{j}$).  
\\
\paragraph{High order numerical schemes}
We use accurate high order numerical methods to predict and obtain high order numerical approximations to the system \eqref{eq27a}-\eqref{eq27b}. For hyperbolic conservation law equations, high order schemes capture smooth regions of the solution and reduce the numerical viscosity at discontinuities, numerical diffusion and also spurious oscillations. In particular, essentially non-oscillatory (ENO) methods were originally designed by \cite{harten1987, harten1987-2} using adaptive stencils to obtain information of smooth regions in the presence of discontinuities. 
The main idea of the ENO scheme is using the {\em smoothest} stencil among several candidates to approximate the fluxes at cell boundaries, thereby obtaining high order accuracy, and overcoming all the difficulties explained above. Based on ENO methods, Ref~\cite{liu94} introduced a new version of weighted essentially non-oscillatory (WENO) methods computing the spatial values as a convex combination of the ENO3 cell averaged parabolas.\\
Here, we apply a one-dimensional fifth-order scheme (WENO5) based on the point-wise ENO3 parabolas that was developed in \cite{jiang96} as an approximation of hyperbolic conservation laws of the equations presented in \eqref{a1}.
For this purpose, we use the Godunov scheme for calculating the numerical fluxes , $\hat{f}_{j+1/2}:=g_{GOD}(u_j^n,u_{j+1}^n)$  and $\hat{f}_{j-1/2}:=g_{GOD}(u_{j-1}^n,u_{j}^n)$ computed at the interface, $x_{j+1/2}$.
We define a computational grid  $x_{j-1/2}\leq x \leq x_{j+1/2}$ and a time discretization where  the computational cells in our system are defined as $C_j^n = \left[x_{j-1/2},x_{j+1/2}\right] \times \left[t_{n},t_{n+1}\right]$. We can use a high order reconstruction procedure for every $j$ giving the numerical approximations of a piecewise smooth function $g(x)$ from its cell averages, \[u_j = \frac{1}{h}\int^{x_{j+1/2}}_{x_{j-1/2}}g(\xi,t_n)d\xi\]We are concerned with the fifth order accurate reconstructions which means that for every $i$, we have an $R_j$ defined on $C_j$  that reconstructs $g(x)$ on $C_j$,
\[g(x) - R_j(x) = O(h^5)\] 
The trasnformation of the reconstructed solution into the reconstructed fluxes from the cell interfaces can be possible due to \citep{shu89}. The numerical fluxes are thus reconstructed from the data of fifth-point variables stencil points $u_{j-2},u_{j-1},u_{j},u_{j+1}$ and $u_{j+2}$ to compute the lateral derivatives $x_{j-\frac{1}{2}}$ and $x_{j+\frac{1}{2}}$. The nonlinear weights for $k = 0,1,2$:
\begin{equation}
w_k=  \frac{a_k}{w_0 + ... + w_{2}},  \quad \textit{for} \quad k = 0,1,2
\end{equation}
where
\begin{equation}
a_0 = \frac{0.1}{(\epsilon + I S_0)^2}, \quad a_1 = \frac{0.6}{(\epsilon + I S_1)^2} \quad \textit{and} \quad a_2 = \frac{0.3}{(\epsilon + I S_2)^2}.
\end{equation}
are based on smoothness indicators,
\begin{eqnarray}
&&IS_0= \frac{13}{12}(u_{j-2}-2u_{j-1}+u_j)^2 + \frac{1}{4}(u_{j-2}-4u_{j-1}+3u_j)^2 \nonumber\\ 
&&IS_1= \frac{13}{12}(u_{j-1}-u_{j}+u_{j+1})^2 + \frac{1}{4}(u_{j-1}-u_{j+1})^2\\
&&IS_2= \frac{13}{12}(u_{j}-2u_{j+1}+u_{j+2})^2 + \frac{1}{4}(3u_{j}-4u_{j+1}+u_{j+2})^2\nonumber
\end{eqnarray}
Here, $\epsilon$ is a positive real number introduced to prevent division by zero, in our numerical simulations $\epsilon = 1.0e-12$.
To compute the numerical fluxes, we will have a nonlinear convex combination of the ENO parabolas, thereby achieving fifth order accuracy in space on smooth regions. For the right and left fluxes respectively
\begin{widetext}
\begin{eqnarray}
&&u_{\text{right}} = w_0 \cdot (\frac{1}{3}u_{j-2} - \frac{7}{6}u_{j-1} + \frac{11}{6}u_j) +
w_1\cdot (-\frac{1}{6}u_{i-1} - \frac{5}{6}u_{j} + \frac{1}{3}u_{j+1})+ w_2 \cdot (\frac{1}{3}u_{j} - \frac{5}{6}u_{j+1} + \frac{1}{6}u_{i+2}),\\ \label{u_l}
&&u_{\text{left}} = w_0 \cdot (-\frac{1}{6}u_{j-2} + \frac{5}{6}u_{j-1} + \frac{1}{3}u_j) +
w_1\cdot (\frac{1}{3}u_{j-1} + \frac{5}{6}u_{j} - \frac{1}{6}u_{j+1})+ w_2 \cdot (\frac{1}{3}u_{j+2} - \frac{7}{6}u_{j+1} + \frac{11}{6}u_{j}). \label{u_r}
\end{eqnarray}\end{widetext}
To summarize, the WENO method has the following ingredients to perform the numerical scheme. It is based on approximating the numerical fluxes using a convex combination of all the candidate stencils. In addition, a weight is associated to each candidate stencil and depends on the contribution of each one.
The procedure to obtain numerical methods of high accuracy is to divide the domain interval of the function into subintervals, and approximate locally the original function using an elementary function (parabolic reconstruction) for each subinterval. The the high-order numerical fluxes at the interface $x_{j+1/2}$ are computed using formulas \eqref{u_l} and \eqref{u_r} to get  \begin{equation}
\hat{f}_{j+1/2} = g_{GOD}(u_{left},u_{right})
\end{equation}
The evolution in time (see \cite{shu88}) is a third-order total variation diminishing (TVD) Runge-Kutta which follows
\begin{equation}
u^{(i)}_j = \sum_{k=0}^{i-1}[\alpha_{ik}\cdot u_j^{(k)} + \beta_{ik}\cdot(-\frac{\Delta t}{h})\cdot(\hat{f}^{(k)}_{j+1/2}-\hat{f}^{(k)}_{j-1/2})], \, i=1,2,3.
\end{equation}
where
\begin{equation}
u_j^{(0)} = u_j^n, \quad \quad u^{(3)}_j = u^{(n+1)}_j 
\end{equation}
and with positive coefficients (see Table~\ref{table:2}).
\begin{table}
\centering
\begin{tabular}{  c c c c c c } 
 \hline
 $\alpha_{1k}$ & $\alpha_{2k}$ & $\alpha_{3k}$& $\beta_{1k}$& $\beta_{2k}$& $\beta_{3k}$ \\ \hline
 $1$ & $0$ & $0$& $1$ &$0$ &$0$ \\ 
 $\frac{3}{4}$ & $\frac{1}{4}$ & $0$& $0$ &$\frac{1}{4}$& $0$\\ 
 $\frac{1}{3}$ & $0$ & $\frac{2}{3}$& $0$ &$0$ &$\frac{2}{3}$ \\
 \hline
\end{tabular}
\caption{Third order TVD Runge-Kutta scheme}\label{table:2}
\end{table}

\end{document}